\documentclass[prl,twocolumn,a4paper,longbibliography]{revtex4-2}
\pdfoutput=1

\usepackage[paperwidth=210mm,paperheight=297mm,centering,hmargin=1.65cm,vmargin=2.04cm]{geometry}

\date{\today}

\newcommand{\kB}{k_{\textrm{B}}}
\newcommand{\Ec}{E_{\textrm{c}}}
\newcommand{\kc}{k_{\textrm{c}}}
\newcommand{\kD}{k_{\textrm{D}}}

\newcommand{\UD}{U_{\textrm{D}}}

\newcommand{\ts}{t_{\textrm{s}}}
\newcommand{\thold}{t_{\textrm{hold}}}
\newcommand{\Us}{U_{\textrm{s}}}
\newcommand{\ws}{\omega_{\textrm{s}}}
\newcommand{\kr}{k_r}

\newcommand{\omegac}{\omega_{\rm peak}}
\newcommand{\Gammad}{\Gamma_{\rm d}}

\usepackage{float}

\usepackage{fourier}

\usepackage{color}
\usepackage{graphicx}
\usepackage{amsmath,amssymb}

\usepackage{times}
\usepackage{upgreek}
\usepackage{psfrag} 
\usepackage{latexsym} 
\usepackage{amstext}
\usepackage{amsxtra} 
\usepackage{soul}
\usepackage{textcomp}
\usepackage{amsfonts}
\usepackage{graphicx}
\usepackage{bm}
\usepackage{color}
\usepackage{braket}
\usepackage{units}
\usepackage[svgnames]{xcolor}
\definecolor{myColor}{rgb}{0.02,0.12,0.3}
\definecolor{myciteColor}{rgb}{0.39,0.7,0.89}
\usepackage[colorlinks=true,citecolor=myColor,linkcolor=myColor,urlcolor=myColor]{hyperref}

\def\be{\begin{equation}}
\def\ee{\end{equation}}

\makeatletter
\def\@fnsymbol#1{\ensuremath{\ifcase#1\or \dagger\or *\or \ddagger\or
   \mathsection\or \mathparagraph\or \|\or **\or \dagger\dagger
   \or \ddagger\ddagger \else\@ctrerr\fi}}
\makeatother

\begin{document} 

\title{
Observation of subdiffusive dynamic scaling in a driven and disordered Bose gas
}

\author{Gevorg Martirosyan}
\author{Christopher J. Ho}
\author{Ji\v{r}\'{i}~Etrych}
\author{Yansheng Zhang}
\author{Alec Cao}
\altaffiliation{Present address: JILA, NIST, and Department of Physics, University of Colorado, Boulder, Colorado 80309, USA}
\author{Zoran Hadzibabic}
\author{Christoph~Eigen}
\email{ce330@cam.ac.uk}
\affiliation{
Cavendish Laboratory, University of Cambridge, J. J. Thomson Avenue, Cambridge CB3 0HE, United Kingdom}

\begin{abstract}
We explore the dynamics of a tuneable box-trapped Bose gas under strong periodic forcing in the presence of weak disorder. In absence of interparticle interactions, the interplay of the drive and disorder results in an isotropic nonthermal momentum distribution that shows subdiffusive dynamic scaling, with sublinear energy growth and the universal scaling function captured well by a compressed exponential. 
We explain that this subdiffusion in momentum space can naturally be understood as a random walk in energy space.
We also experimentally show that for increasing interaction strength, the gas behavior smoothly crosses over to wave turbulence characterized by a power-law momentum distribution, which opens new possibilities for systematic studies of the interplay of disorder and interactions in driven quantum systems.
\end{abstract}
\maketitle

Complex microscopic behavior of both classical and quantum systems can often be characterized by universal statistical properties. While such descriptions are more commonly associated with thermodynamic equilibrium, far-from-equilibrium systems, from kicked rotors and chaotic billiards~\cite{Fishman:1982,Bohigas:1984,Lichtenberg:1992,Stockmann:1999} to turbulent fluids~\cite{Kolmogorov:1941, Obukhov:1941, Zakharov:1992}, can also display emergent universal behavior. 
A fascinating manifestation of this is dynamic scaling, which is akin to the scale-invariance of equilibrium systems close to a phase transition, but generalized to scaling in both space and time. Such behavior is known from surface growth~\cite{Family:1985,Kardar:1986} and both normal and anomalous diffusion ~\cite{Shlesinger:1993, Metzler:2000}. Recently, dynamic scaling was observed in a variety of quantum systems and in different scenarios~\cite{Sagi:2012,Pruefer:2018, Erne:2018, Bouganne:2020, Glidden:2021, Galka:2022,Wei:2022, Joshi:2022, Fontaine:2022, Orozco:2022, Huh:2023, Lannig:2023, Gazo:2023}, including the relaxation of atomic gases~\cite{Erne:2018, Pruefer:2018,Glidden:2021, Wei:2022, Huh:2023, Lannig:2023, Orozco:2022, Gazo:2023} and polariton systems~\cite{Fontaine:2022}, and the build-up of wave turbulence in a driven  interacting Bose gas~\cite{Galka:2022}. These experiments provide mounting evidence for the hypothesis that such scaling is generic to far-from-equilibrium quantum systems~\cite{Berges:2008}.

Usually interactions are at the heart of the emergent dynamics, but naturally present disorder can also play a crucial role.
The study of disorder is a vast field, with highlights including localization and quantum-Hall phenomena in 2D electron gases and quantum wires~\cite{Anderson:1958a,Von-Klitzing:1986, Altshuler:1980,Evers:2008,Castro-Neto:2009}, coherent backscattering of acoustic and electromagnetic waves~\cite{Sheng:1990,Labeyrie:1999}, and Anderson localization of cold atoms~\cite{Billy:2008, Roati:2008}.
Moreover, the interplay of disorder and interactions can result in new phenomena such as many-body localization in lattice systems~\cite{Nandkishore:2015, Schreiber:2015, Abanin:2019} and time crystals~\cite{Zhang:2017, Choi:2017, Else:2020, Kongkhambut:2022}.

\begin{figure}[b!]
\centerline{\includegraphics[width=1\columnwidth]{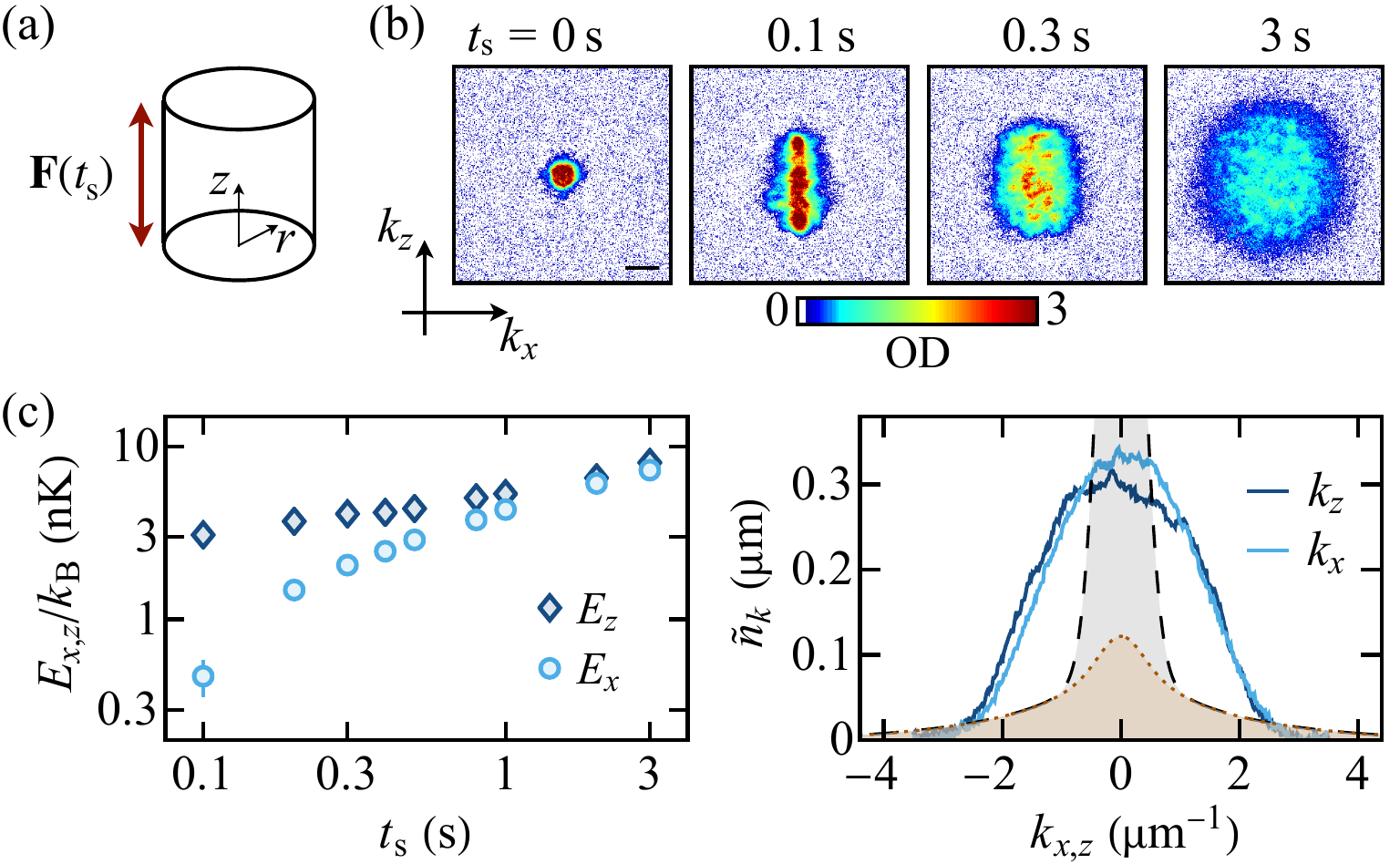}}
\caption{
Noninteracting box-trapped Bose gas driven far from equilibrium.
(a)~Box geometry and driving force, ${\bf F}=(\Us/L) \cos(\ws \ts)\, {\bf \hat{z}}$, where $L\approx50\,\upmu$m is the box length.
(b)~Time-of-flight images, giving 2D momentum distributions $n_k(k_x,k_z)$, for \mbox{$N=3.3 \times 10^5$} atoms in a box of depth $\UD/\kB=90$\,nK, driven at $\ws/(2\pi)=10$\,Hz with $\Us/\kB=10.5$\,nK. The scale bar shows $1\,\upmu$m$^{-1}$ and the optical density (OD) saturates at $3$.
(c)~Energies, $E_{x,z}$, and 1D momentum distributions, $\tilde{n}_{k}(k_{x,z})$, obtained by integrating 2D distributions over $k_z$ or $k_x$; note that $E_x \approx E_z$ at $\ts =0$. The axial $E_z$ initially (in $< 0.1\,$s) rises far above $E_x$, but at long $\ts$ the two energies are almost equal and grow in unison.
The long-time momentum distribution is essentially isotropic [$\tilde{n}_{k}(k_{x}) \approx \tilde{n}_{k}(k_{z})$], but highly nonthermal. We show $\tilde{n}_{k}(k_{x,z})$ for $\ts = 3\,$s, together with the calculated equilibrium distribution for the same energy per particle, $E/\kB\approx 23$\,nK (shaded curve); the experimental distributions show no condensate peak, even though the condensation temperature is $180$\,nK and the equilibrium distribution has $66\%$ condensed fraction (gray)~\cite{ConvFootnote}. 
}
\label{fig1a}
\end{figure}

In this Letter, we explore the dynamics of a 3D box-trapped Bose gas strongly driven in presence of weak disorder. In absence of interatomic interactions the gas shows subdiffusive dynamic scaling: its energy grows sublinearly with the drive time $\ts$, approximately as $\ts^{0.5}$, and its momentum distribution at different $\ts$ is described by a scaling function that is captured well by an isotropic compressed exponential.
This behavior is in stark contrast to that expected for a disorder-free noninteracting gas in our cylindrical geometry with forcing along the box axis [see Fig.~\ref{fig1a}(a)]; in that case the system is effectively 1D and one expects chaotic dynamics with bounded energy growth~\cite{Supplementary,Reichl:1986,Lin:1988}.
Our observations can be explained in terms of a random walk in energy space (see~\cite{YZhang:2023} for our detailed theoretical study) and give credence to the proposals that such random walks are a generic feature of thermally isolated driven systems~\cite{Jarzynski:1993,Bunin:2011, Hodson:2021}.
We also experimentally show, by tuning the interaction strength, that the energy-space random walk observed in the noninteracting limit is continuously connected to wave turbulence characterized by a power-law scaling function~\cite{Navon:2016, Navon:2019, Galka:2022, Dogra:2023}.
This points to interesting future studies in the regime where the drive, the disorder, and the interactions all play a significant role.

\begin{figure}[t!]
\centerline{\includegraphics[width=1\columnwidth]{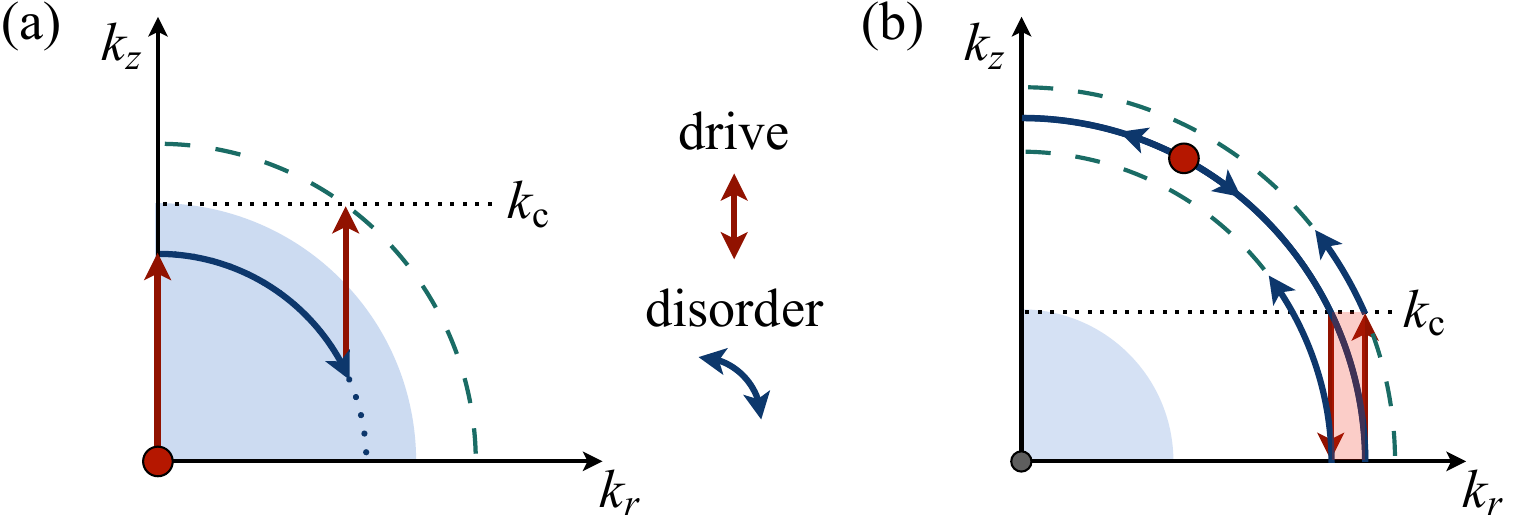}}
\caption{
Semiclassical picture of a driven and disordered noninteracting Bose gas.
(a)~The drive mixes only states with $k_z$ values up to some $\kc$~\cite{Supplementary}, so in absence of disorder the radial momentum $k_r$ (in the $k_x - k_y$ plane) remains zero, and the growth of a particle's energy is bounded by $\Ec = \hbar^2 \kc^2/(2m)$.
However, disorder-induced elastic scattering distributes energy into the radial modes, and allows the drive to populate states above $\Ec$ (outside the blue shaded area).
(b)~This picture implies subdiffusive long-time dynamics, with a sublinear energy growth.
Here we consider a particle (red dot) that already has an energy above $\Ec$, and $k_z > \kc$, so it does not interact with the drive. If elastic scattering (solid circle) reduces its $k_z$ to below $\kc$, the particle temporarily interacts with the drive, until another scattering event increases its $k_z$ above $\kc$. The interaction with the drive can either increase or decrease the particle's energy, as exemplified by the red arrows. The alternation of scattering and driving events thus results in an energy-space random walk, with a characteristic step size $\Ec$.
}
\label{fig1b}
\end{figure}

We start with a quasi-pure $^{39}$K Bose--Einstein condensate (BEC) in the lowest hyperfine state, trapped in a cylindrical optical box~\cite{Gaunt:2013,Eigen:2016,Navon:2021}. The condensate is prepared at a scattering length $a=200\,a_0$ (where $a_0$ is the Bohr radius), and we slowly (in $5$\,s) tune $a$ to zero by tuning the bias magnetic field to $350.4(1)$\,G~\cite{Fattori:2008b,Etrych:2023}.
For a noninteracting BEC in our box of length $L \approx 50 \,\upmu$m and radius $R \approx 15 \,\upmu$m, the frequency of the lowest-lying axial excitation is (ignoring disorder) $\omega_z=3 \pi^2 \hbar /(2mL^2)\approx 2\pi \times 1.5 $\,Hz, where $m$ is the atom mass, while our variable trap depth $\UD$ is always larger than $2\pi\hbar \times 400$\,Hz.
Weak optical disorder, proportional to the trapping laser power and hence $\UD$,  is always present in our holographically created trap~\cite{Gaunt:2012, Gaunt:2013, Gaunt:2014-th}, but is typically irrelevant in interacting-gas experiments~\cite{Gotlibovych:2014}.
We inject energy into the system along the box axis $\bf{z}$, using a spatially uniform time-varying force~\cite{Navon:2016} of magnitude $F(\ts)=(\Us/L) \cos(\ws \ts)$, with $\ws > \omega_z$ and $\hbar \omega_z\ll\Us\ll \UD$ (see also~\cite{Supplementary}).
After driving the gas for a variable time $\ts$, we probe its momentum distribution using absorption imaging after $50$\,ms of time-of-flight expansion~\cite{ResFootnote}; this gives line-of-sight integrated 2D distributions $n_k ({\bf k})$, which we normalize such that $\int n_k({\bf k})\, {\rm d}{\bf k} = 1$.

In Figs.~\ref{fig1a}(b,c) we illustrate our qualitative observations; here $E_{x,z} = \int \hbar^2 k_{x,z}^2/(2m) n_k(k_x, k_z) \, {\rm d} {\bf k}$ and 1D momentum distributions, $\tilde{n}_{k}(k_{x,z})$, are obtained by integrating $n_k(\bf{k})$ over $k_z$ or $k_x$. Initially the dynamics is essentially 1D, with the drive rapidly (in $< 0.1\,$s) increasing only $E_z$. This is what is expected in absence of disorder, in which case the growth of $E_z$ would be bounded~\cite{Supplementary}. However, at long times $E_x \approx E_z$ and the energy keeps growing. The long-time  momentum distributions along $k_x$ and $k_z$  are nearly identical, but far from thermal; they show no BEC peak even though the energy per particle is far below the equilibrium condensation value.

In Fig.~\ref{fig1b}(a) we outline a semiclassical picture of how the interplay of the drive and disorder can lead to isotropic dynamics with unbounded energy growth. 
The drive mixes only axial modes, with $k_z$ up to some $\kc$~\cite{Supplementary}. 
The disorder-induced elastic scattering transfers energy into the radial modes and allows the drive to increase a particle's energy above $\Ec = \hbar^2 \kc^2/(2m)$. Alternating scattering and driving events then lead to an unbounded energy growth. 

In Fig.~\ref{fig1b}(b) we explain why at long times this growth is sublinear, corresponding to subdiffusion in momentum space. Once the average energy per particle, $E$, is significantly larger than $\Ec$, most particles have $k_z > \kc$ and do not interact with the drive. When a particle is occasionally scattered in and out of the $k_z < \kc$ space, its temporary interaction with the drive can either increase or decrease its energy (red arrows).
This results in an energy-space random walk with a characteristic step size $\Ec$~\cite{YZhang:2023}:
\begin{equation}
    \frac{\text{d}}{\text{d} \ts} \left( E^2 \right) \propto r(E) \, \Ec^2 \, .
    \label{eq:qualitative-random-walk}
\end{equation}
The rate $r$ is energy-dependent, because at any time only a fraction of particles $\propto \hbar \kc/\sqrt{2mE}$ interacts with the drive, and because the density of states for elastic scattering is $\propto \sqrt{E}$.
As shown in Ref.~\cite{YZhang:2023}, this model predicts $E \propto \ts^{\eta}$, with $2/5 \leq \eta \leq 1/2$ depending on the ratio of the elastic scattering rate and the rate at which the drive mixes $k_z$ states.

\begin{figure}[t!]
\centerline{\includegraphics[width=\columnwidth]{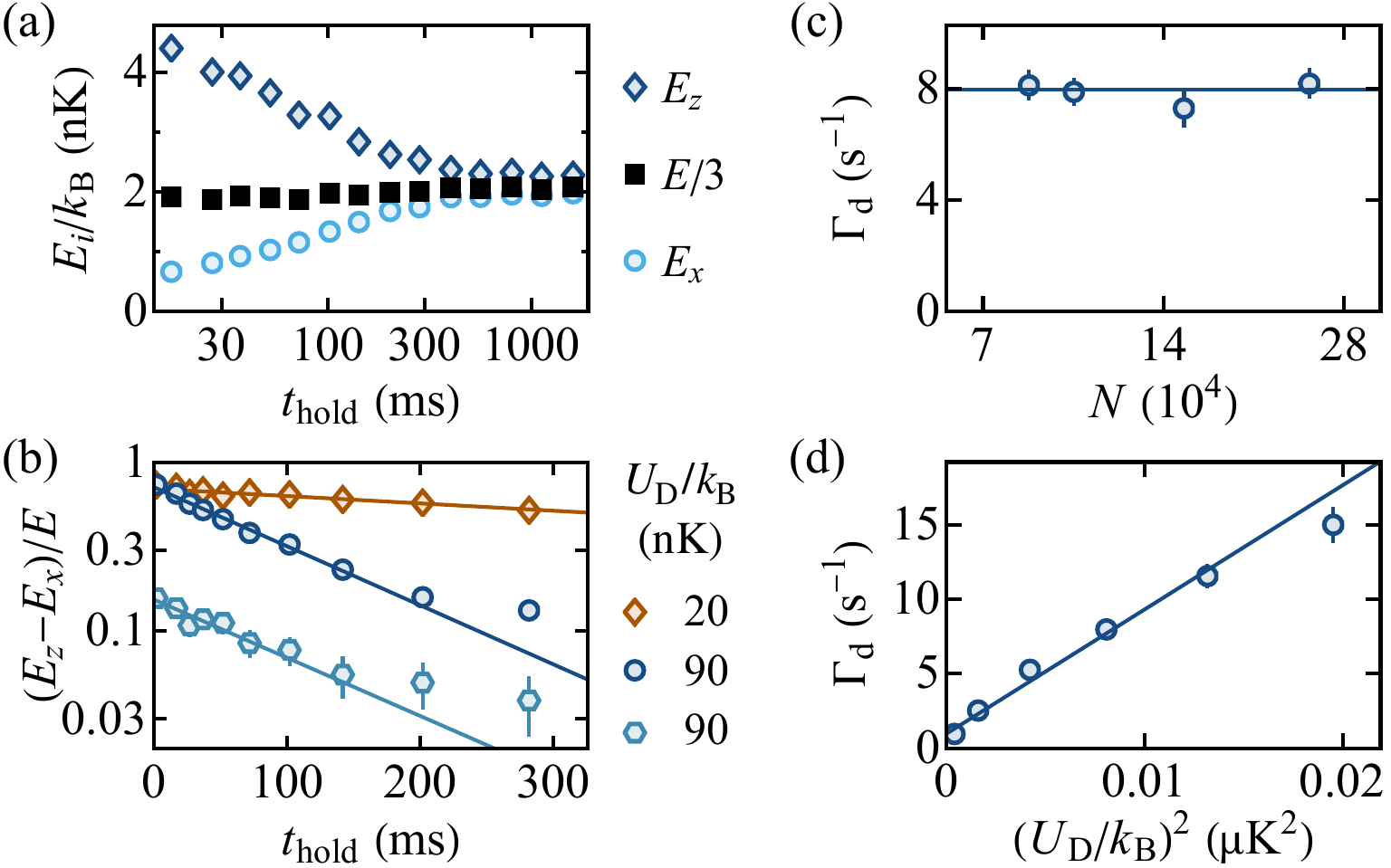}}
\caption{
Disorder-induced cross-dimensional coupling.
(a) Here we stop the drive with $\Us/\kB=10.5\,$nK and $\ws/(2\pi)= 10$\,Hz after $\ts=0.1$\,s [see Fig.~\ref{fig1a}(b)] and then hold the gas for a variable time $\thold$ in a trap with $\UD/\kB=90$\,nK. 
The energies $E_{z}$ and $E_{x}$ both relax towards $E/3 = (E_z+2E_x)/3$ (by symmetry $E_x = E_y$).
(b)~The decay of the anisotropy $(E_z - E_x)/E$ is initially exponential, \mbox{$
\propto \exp(-\Gammad\thold)$} (solid lines); $\Gammad$ grows with disorder strength ($\propto \UD$) and is independent of the initial anisotropy.
(c)~As expected for single-particle scattering, $\Gammad$ is independent of the gas density ($\propto N$); here $\UD/\kB=90$\,nK.
(d)~Dependence of $\Gammad$ on $\UD$. 
Up to a small offset of $1\,$s$^{-1}$, the data are captured by our numerical simulations with rms disorder equal to $2$\% of $\UD$ (solid line).
}
\label{fig2}
\end{figure}

\begin{figure*}[t!]
\centerline{\includegraphics[width=1\textwidth]{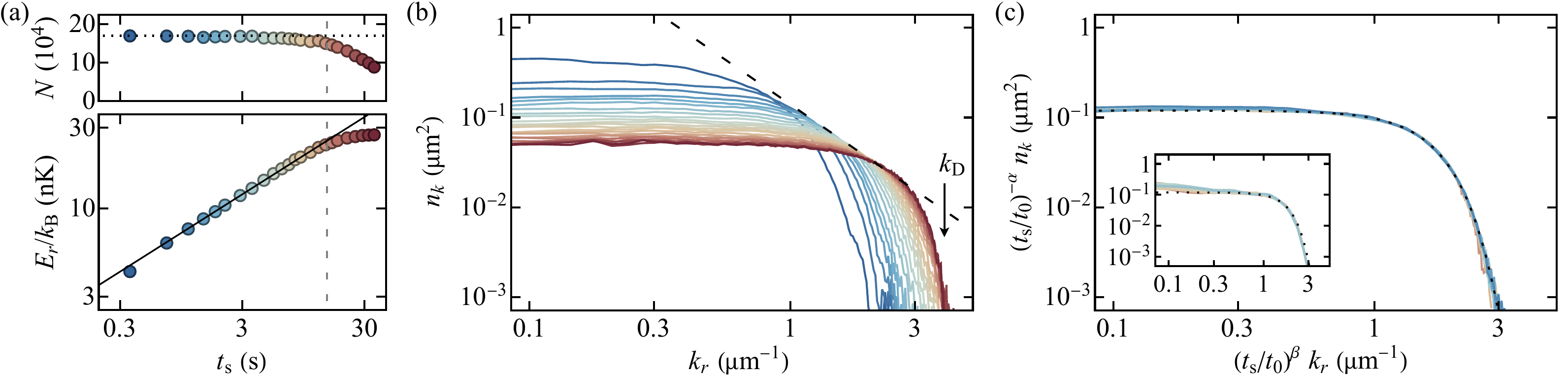}}
\caption{
Subdiffusive dynamic scaling; here $\Us/\kB=7.0$\,nK, $\ws/(2\pi)=10$\,Hz, and $\Gammad=8.0\,\text{s}^{-1}$.
(a) Evolution of the atom number $N$ and the radial per-particle energy $E_r$. For $\ts \lesssim 15$\,s (vertical dashed line), $N$ is essentially constant~\cite{NFootnote}, while $E_r \propto \ts^\eta$ with $\eta = 0.46$ (solid line); at longer $\ts$ some particles have enough energy to leave the trap, so $N$ drops and $E_r$ saturates.
(b) Evolution of the radial momentum distribution $n_k(k_r)$; each curve corresponds to a point in (a), with the same color coding. The dashed line, $\propto k_r^{-2}$, is tangent to all the self-similar curves, and $\kD$ is the momentum-space trap depth.
(c) For scaling exponents $\alpha=-0.45$ and $\beta=-0.23$ (and arbitrarily chosen reference time $t_0=3$\,s), the distributions for $\ts\in [1.1,14.4]$\,s and all $k_r$ collapse onto a universal curve.
The dotted line shows a compressed-exponential fit, $f_{\rm ce} \propto \exp{\left[-\left(k_r/k_0\right)^\kappa\right]}$, with $\kappa=3.0$ (see text).
The inset shows the results of numerical simulations with the same drive, disorder, and scaling parameters (see text and \cite{Supplementary}); for comparison with the experiments, the dotted line is the same as in the main panel.
}
\label{fig3}
\end{figure*}

To isolate and quantify the disorder-induced scattering in our system, we prepare an anisotropic $n_k$ with a short $\ts$, stop the drive, and study the subsequent cross-dimensional relaxation for different disorder strengths ($\propto \UD$).
In Fig.~\ref{fig2}(a) we show an example of how $E_z$ and $E_x$ both approach $E/3$, and in Fig.~\ref{fig2}(b) we show how the anisotropy $(E_z - E_x)/E$ decays for different $\UD$. The initial decay is captured well by an exponential (solid lines), and we use the decay constant $\Gammad$ as a measure of the typical scattering rate~\cite{kFootnote}.
At long times, the anisotropy decay slows down, which we also observe in simulations of the Schr\"{o}dinger equation with disorder~\cite{Supplementary}, and attribute to the quantization of states in a finite-size box~\cite{AniFootnote}.
In Fig.~\ref{fig2}(c) we show that, as expected for single-particle scattering, $\Gammad$ is independent of the particle density.

In Fig.~\ref{fig2}(d) we show the dependence of $\Gammad$ on $\UD$. The solid line is based on our numerical simulations~\cite{Supplementary}, which give $\Gammad \propto \UD^2$, as expected from perturbation theory. We match the data well by setting the rms disorder strength to $2$\% of $\UD$, and adding a small offset to $\Gammad$. 
The $2$\% disorder is compatible with $\approx 1\%$ observed in the bench tests of the optical potentials used for our box trap~\cite{Gaunt:2014-th} (see also \cite{Gaunt:2012,Schroff:2023}), and also with the measurements with atoms in Ref.~\cite{Gotlibovych:2014}, where the uniformity of the gas density was confirmed down to energies of a few \% of $\UD$. In simulations we assume that the disorder is uncorrelated down to a lengthscale of $800\,$nm, set by the simulation grid, which is comparable with the expected correlation length of experimental disorder, set by the trap-laser wavelength, $\lambda = 532\,$nm. The small offset in $\Gammad$ could arise from trap-shape imperfections or magnetic-field inhomogeneity.

We now turn to the study of the long-time isotropic dynamics for continuous driving (Fig.~\ref{fig3}). Here we take images along the drive axis ${\bf z}$, which avoids the small effects of the center-of-mass oscillation~\cite{Supplementary}. The distribution in the $k_x-k_y$ plane is always isotropic, $ n_k(k_x, k_y) = n_k(k_r)$, where $k_r=(k_x^2+k_y^2)^{1/2}$. We normalize $\int 2\pi k_r n_k(k_r)\, {\rm d}k_r = 1$ and define $E_r = E_x + E_y$, so $E_r \approx 2E/3$ for long $\ts$. 

In Fig.~\ref{fig3}(a) we plot $N(\ts)$ and $E_r(\ts)$ for $\Us/\kB=7.0$\,nK, $\ws/(2\pi) = 10$\,Hz, and \mbox{$\Gammad=8.0$\,s$^{-1}$}. At $\ts \approx 15$\,s some atoms reach the momentum-space trap depth $\kD = \sqrt{2m\UD/\hbar^2} = 3.8\,\upmu \text{m}^{-1}$, at which point $N$ starts dropping and $E_r$ saturates. Until then, $N$ is essentially constant~\cite{NFootnote} and $E_r$ shows power-law growth, $E_r\propto \ts^{\eta}$, with $\eta=0.46(2)$.

In Fig.~\ref{fig3}(b) we show the evolution of $n_k(k_r)$. For \mbox{$\ts \gtrsim 1$\,s} the distributions are self-similar, with a well-defined front moving towards the UV until it reaches $\kD$; for $\ts\gtrsim15$\,s the distribution is essentially stationary. 

The fact that $n_k$ is self-similar for $1\,{\rm s} \lesssim \ts \lesssim 15\,{\rm s}$, while $E_r$ grows algebraically, implies dynamic scaling:
\begin{equation}
   n_k(k_r, \ts)=\big{(}\tfrac{\ts}{t_0}\big{)}^{\alpha}\,n_k \left( \big{(} \tfrac{\ts}{t_0} \big{)}^{\beta}k_r,t_0 \right) ,
   \label{eq:scaling}
\end{equation}
with $\beta = -\eta/2$, reflecting the subdiffusive energy growth, and $\alpha = 2\beta$, reflecting a particle-conserving transport; $t_0$ is an arbitrary reference time.

In Fig.~\ref{fig3}(c) we show that, for $\ts\in [1.1,14.4]$\,s and all $k_r$, the distributions can be collapsed onto a universal curve, with $\alpha=-0.45(2)$ and $\beta=-0.23(1)$~\cite{Supplementary}.
The calculations in~\cite{YZhang:2023} predict such scaling with the 3D momentum distribution captured by a compressed exponential, \mbox{$\propto\exp{\left[-\left(k/k_0'\right)^{\kappa_{\rm 3D}}\right]}$}, with $\kappa_{\rm 3D}$ varying between $4$ (for $\eta = 1/2$) and $5$ (for $\eta = 2/5$), and $k_0' \propto \ts^{1/\kappa_{\rm 3D}}$~\cite{ExpFootnote}.
We empirically find that $n_k(k_r)$, obtained by integrating the 3D distribution along $k_z$, is captured by a normalized compressed exponential $f_{\rm ce} = \left[\pi k_0^2\, \Gamma(1+2/\kappa)\right]^{-1} \exp{\left[-\left(k_r/k_0\right)^\kappa\right]}$ with a reduced $\kappa=3.0(2)$ [dotted line in Fig.~\ref{fig3}(c)]~\footnote{Integrating a 3D compressed exponential along $k_z$ does not analytically give another compressed exponential, but this is a good approximation in the experimental $k$ range.}.
As shown in the inset of Fig.~\ref{fig3}(c), we reproduce our observations in numerical simulations; here we show the results of simulations for $\ts\in \{2.9-18\}$\,s, obtained with the same $\Us$, $\ws$, and $\Gammad$, and collapsed with the same $\alpha$, $\beta$, and $t_0$ as in the experiments.

Repeating our experiments with various drive and disorder parameters, for $\Us/\kB \in$~[$3.5$,\,$10.5$]\,nK,
$\ws/(2\pi)\in$~[$5$,\,$15$]\,Hz,
and $\Gammad\in$~[$2.5$,\,$15$]\,${\rm s}^{-1}$, we robustly observe dynamic scaling 
with $\eta=0.46(2)$, $\alpha = -0.47(4)$, $\beta=-0.24(2)$, and $\kappa = 2.9(2)$. 
For our parameters, $\eta$ is indeed expected to be in the broad crossover from $1/2$ to $2/5$~\cite{YZhang:2023}.

\begin{figure}[t!]
\centerline{\includegraphics[width=\columnwidth]{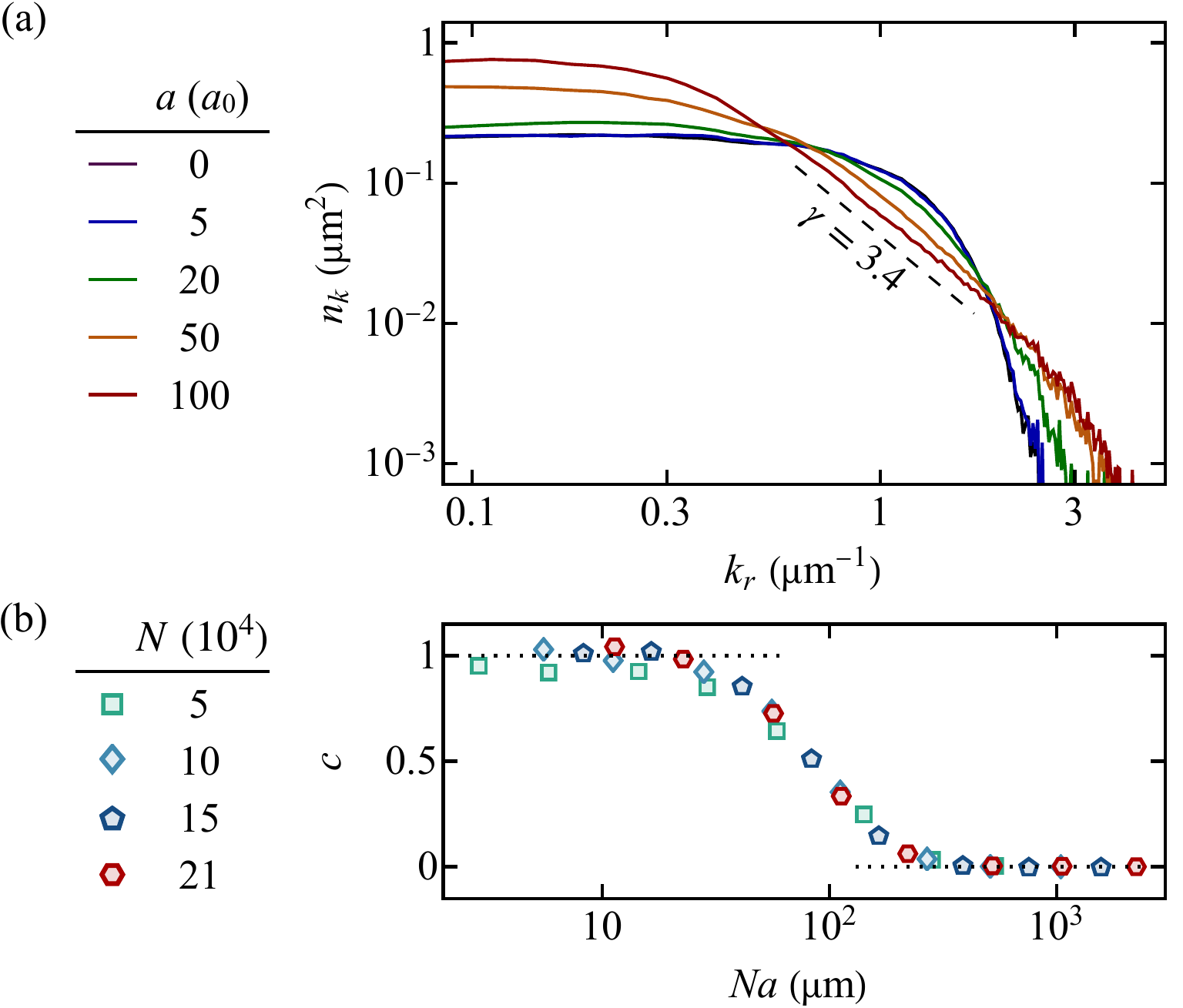}}
\caption{
Crossover from subdiffusion to turbulence, for $\Us/\kB=7.0$\,nK, $\ws/(2\pi) = 10$\,Hz, $\Gammad=15\,{\rm s}^{-1}$, and $\ts=1$\,s. (a) $n_k$ for $N = 10^5$ atoms and different scattering lengths $a$. 
For subdiffusion, $n_k$ is captured by a compressed exponential $f_\mathrm{ce}$. For a turbulent cascade, 
$n_k \propto k_r^{-\gamma +1}$ (dashed line) for $k_r \gtrsim 1/\xi$, where $\xi$ is the healing length; here $\xi=0.6$\,$\upmu\textrm{m}^{-1}$ for $100\,a_0$.
(b)~We quantify the crossover between the two regimes by fitting $n_k=c f_\mathrm{ce}(k_r) + n_0 k_r^{-\gamma+1}$ (with $\gamma=3.4$) for $0.6\,\upmu {\rm m}^{-1}<k_r<2.5\,\upmu {\rm m}^{-1}$, with $c$ and $n_0$ as free parameters.
For various $N$ and $a$, the crossover from $c=1$ (subdiffusion) to $c=0$ (turbulence) depends only on the product $Na$.
}
\label{fig4}
\end{figure}

We conclude by pointing to an interesting question for future study – what happens if the drive, the disorder, and the interactions all play a significant role? In the noninteracting ($a=0$) dynamics observed here, the rate at which energy is absorbed from the drive decays as $\ts^{\eta – 1}$.
On the other hand, in interacting wave-turbulent cascades~\cite{Navon:2016, Navon:2019, Galka:2022, Dogra:2023} this rate is constant and the turbulent steady state is characterized by $n_k (k_r)\propto k_r^{-\gamma +1}$, with $\gamma=3.3(3)$.
The two types of dynamics are qualitatively different, but are continuously connected by tuning $a$, as we illustrate in Fig.~\ref{fig4}(a) for one set of drive and disorder parameters, and fixed gas density ($\propto N$). 
This crossover should be controlled by some dimensionless parameter(s), but from the drive, disorder, and interaction properties, one can construct many candidates. Moreover, further qualitatively different outcomes are possible: for strong-enough interactions, thermalization should be the fastest process, while for strong-enough disorder, localization effects should prevail. Constructing the full dynamical phase diagram for the driven and disordered interacting gas is thus a fascinating challenge. 
As a first step in this direction,
in Fig.~\ref{fig4}(b) we show that for our parameters the crossover from an energy-space random walk to turbulence depends on the product $Na$, which suggests that it can be captured within the mean-field Gross--Pitaevskii framework.

In summary, we have observed subdiffusive dynamic scaling in a noninteracting Bose gas driven far from equilibrium in the presence of weak disorder, which we explain in terms of an energy-space random walk. 
The tuneability of our system opens the possibility to study the interplay of the drive, disorder, and interactions in regimes where they all play a significant role, which we illustrate by showing how the energy-space random walk crosses over to turbulent-cascade dynamics.
Our far-from-equilibrium states with low and tuneable energy per particle also provide a novel starting point for studies of equilibration in closed quantum systems~\cite{Berges:2008,Chantesana:2019,Gazo:2023}.

We thank Nigel R.~Cooper, Nir Navon, Maxim Olshanii, and Theo Geisel for discussions, and Robert P. Smith, Jake~A.~P. Glidden,  Lena H.~Dogra, Timon A. Hilker, and Samuel J. Garratt for early contributions.
This work was supported by EPSRC [Grants No.~EP/N011759/1 and No.~EP/P009565/1], ERC (UniFlat), and STFC [Grant No.~ST/T006056/1].
A.~C. acknowledges support from the NSF Graduate Research Fellowship Program (Grant No. DGE2040434).
Z.~H. acknowledges support from the Royal Society Wolfson Fellowship.
C.~E. acknowledges support from Jesus College (Cambridge). 


\begin{thebibliography}{68}
\bibitem [{Con()}]{ConvFootnote}%
  \BibitemOpen
  \href@noop {} {}\bibinfo {note} {We convolve the equilibrium Bose
  distribution with a Gaussian of width $0.3\,\upmu{\rm m}^{-1}$ (our $k$-space
  resolution~\cite{ResFootnote}).}\BibitemShut {Stop}%
\bibitem [{Res()}]{ResFootnote}%
  \BibitemOpen
  \href@noop {} {}\bibinfo {note} {Our $k$-space resolution, given by the box
  size and the $50$-ms time-of-flight duration, is $\approx 0.3\,\upmu{\rm
  m}^{-1}$}\BibitemShut {NoStop}%
\bibitem [{kFo()}]{kFootnote}%
  \BibitemOpen
  \href@noop {} {}\bibinfo {note} {Note that in reality the $k$-dependent
  scattering rate is different for different particles.}\BibitemShut {Stop}%
\bibitem [{Ani()}]{AniFootnote}%
  \BibitemOpen
  \href@noop {} {}\bibinfo {note} {The semiclassical picture in
  Fig.~\ref{fig1b} assumes a continuum of states. In reality, states are
  quantized and `elastic' scattering is possible only if there is a final state
  with energy within $\approx \hbar \Gammad$ of the initial one. In our weak
  disorder, $\hbar \Gammad$ is never much larger than the energy-quantization
  scale $\hbar^2 \pi^2 /(2m L^2) = h \times 0.5\,$Hz, so the scattering may not
  be fully isotropic.}\BibitemShut {Stop}%
\bibitem [{NFo()}]{NFootnote}%
  \BibitemOpen
  \href@noop {} {}\bibinfo {note} {Without a drive, $N$ decays very slowly,
  with a $1$-body lifetime of $140\,$s, due to collisions with the background
  vapor in the vacuum chamber. On average, these losses do not change
  $E$.}\BibitemShut {Stop}%
\bibitem [{Exp()}]{ExpFootnote}%
  \BibitemOpen
  \href@noop {} {}\bibinfo {note} {Also note that for the dynamic scaling of
  the 3D distribution $\alpha = 3 \beta$, with the same $\beta$ as for
  $n_k(k_r)$.}\BibitemShut {Stop}%
\end{thebibliography}

%

\setcounter{figure}{0} 
\setcounter{equation}{0} 

\renewcommand\theequation{S\arabic{equation}} 
\renewcommand\thefigure{S\arabic{figure}}


\onecolumngrid
\section{\textsc{Supplemental Material}}

\twocolumngrid
\subsection{\textsc{I. Numerical Simulations}}

\vspace{-1em}

\begin{figure}[b!]
\centerline{\includegraphics[width=\columnwidth]{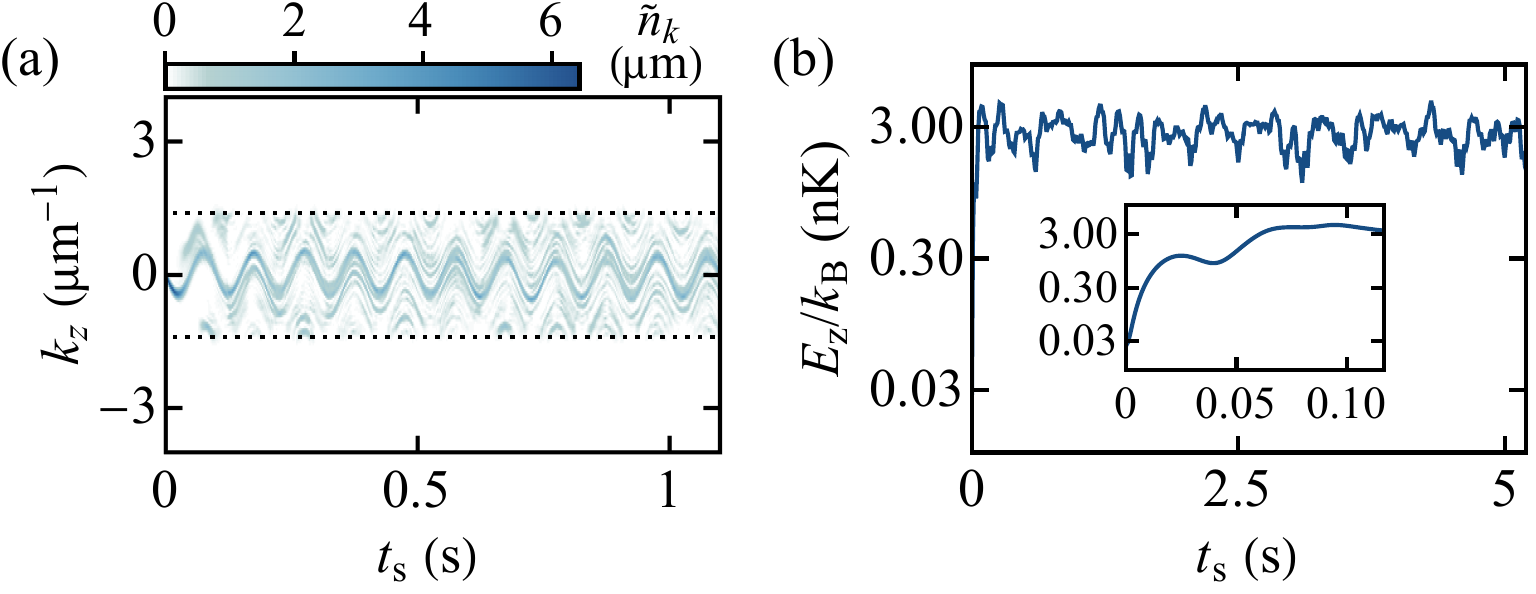}}
\caption{
Simulations of a strongly driven Bose gas in a disorder-free box. (a)~The 1D $\tilde{n}_{k}(k_z)$ versus $\ts$ for $\Us/\kB=10.5$\,nK and $\ws/(2\pi)=10$\,Hz. The dashed lines indicate $\kc$. (b)~The corresponding bounded energy growth. The energy $E_z$ shows large fluctuations around an average value that quickly saturates (see inset).
}
\label{figS1}
\end{figure}

We describe our system using the time-dependent Schr{\"o}dinger equation for a particle trapped in a cylindrical box of radius $R=15\,\upmu{\rm m}$ and length $L=50\,\upmu{\rm m}$ and driven with a potential $V_{\rm s}({\bf r},t) =  \Us  z/L \times{\rm cos }(\omega_{\rm s} \ts)$. We model the disorder present in our system as
$V_{\rm d}(r_i) = \sqrt{3}V_0 X_i$, where $r_i$ are the lattice points in our simulation grid, $X_i$ are independent random numbers sampled uniformly between $-1$ and $1$, and $V_0$ is the rms disorder strength. Our simulations always begin with the system in the ground state at $\ts=0$.

For $V_{\rm 0}=0$, the problem is separable into radial and axial directions, since the drive is oriented along the box axis $\bf {z}$. 
We solve the resultant 1D problem by expanding the wavefunction in the box eigenbasis (using the lowest $80$ states) and solving the coupled differential equations for the expansion coefficients.
In Fig.~\ref{figS1} we plot the calculated momentum distribution $\tilde{n}_k(k_z)$ and energy $E_z$ versus excitation time $\ts$ for $\Us/\kB=10.5$\,nK and $\ws/(2\pi)=10$\,Hz (cf. Fig.~1 in the main text).
The drive only couples $k_z$ modes up to a characteristic momentum $\kc$ [Fig.~\ref{figS1}(a)], and the average $E_z$ saturates (with large fluctuations) after a rapid initial increase [Fig.~\ref{figS1}(b)].

To model the 3D dynamics observed in experiments, we solve the full 3D Schr{\"o}dinger equation using a pseudo-spectral method with fourth-order Runge--Kutta time evolution with a $10\,\upmu{\rm s}$ timestep and a numerical grid of size $100\times 100\times 100\,\upmu$m$^3$ discretized with $\zeta=0.8\,\upmu$m resolution, which is on the order of the trapping laser wavelength, $\lambda=532\,$nm. This results in Fourier components of the disorder up to $\pi/\zeta\approx4\,\upmu{\rm m}^{-1}$, similar to the largest $k$ in the experiment, set by $\kD$.

Simulating the experimental protocol from Fig.~3 of the main text, we find that $V_0/\kB \approx2$\,nK reproduces the experimental $\Gammad=8.0$\,s$^{-1}$ for $\UD/\kB=90$\,nK. In these simulations we also observe, as in the experiments, that the decay of anisotropy slows down at long times.

For the simulations in the inset of Fig.~4(c) of the main text, we have also extracted the scaling exponents as in the experiment (see Section~III) and found $\eta=0.46(1)$,
$\kappa=3.2(1)$ [$\kappa_{\rm 3D}\approx 4$], $\alpha=-0.46(1)$, and $\beta=-0.24(1)$.

\subsection{\textsc{II.~Frequency response}}

\begin{figure}[t!]
\centerline{\includegraphics[width=\columnwidth]{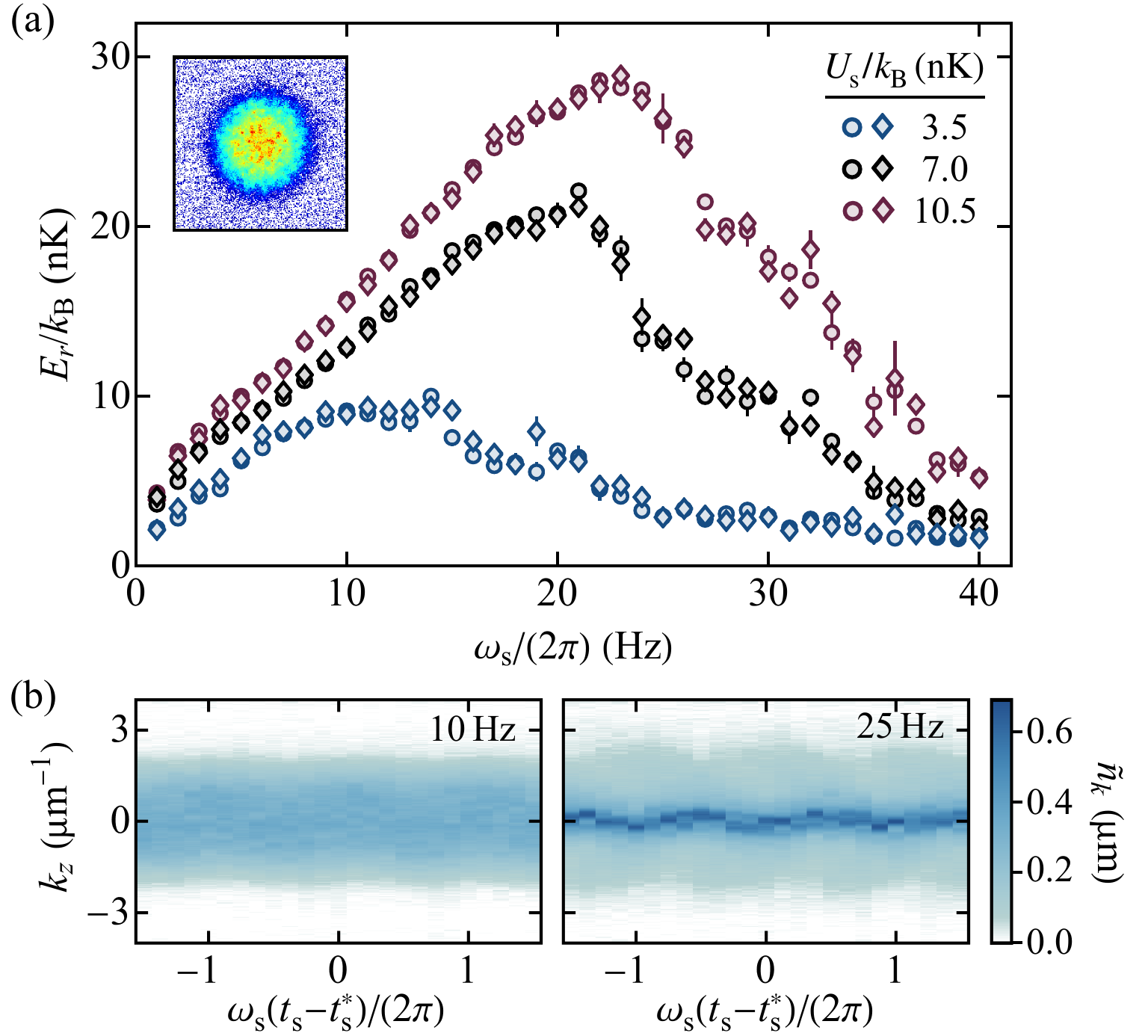}}
\caption{
Frequency response of a noninteracting Bose gas. (a)~$E_r$ versus $\ws$ for a $\ts=4$\,s sinusoidal drive with different $\Us$ (see legend) and $\Gammad=8.0$\,s$^{-1}$. We show data taken with two different initial atom numbers (circles: $2.8\times 10^5$, diamonds: $1.1\times 10^5$).
The inset shows a typical time-of-flight image taken along $\hat{\bf{z}}$ for $\ws<\omegac$.
(b) Evolution of the axial momentum distribution around $\ts^*=4$\,s for $\ws<\omegac$ (left) and $\ws>\omegac$ (right).
}
\label{figS2}
\end{figure}

Here we experimentally study the frequency response of our noninteracting box-trapped gas. 
In Fig.~\ref{figS2}(a), we show for a fixed $\Gammad$ and three different $\Us$ how $E_r$ depends on $\ws$ for a $\ts = 4\,$s sinusoidal drive. At low frequency $E_r$ increases roughly linearly with $\ws$ but above some $\Us$-dependent $\omegac$ the system stops responding strongly.

As illustrated in Fig.~\ref{figS2}(b), we observe qualitatively different behavior below and above $\omegac$. Here we plot the evolution of $\tilde{n
}_k(k_z)$ around $\ts^* = 4$\,s, for $\Us/\kB=7.0$\,nK.
For $\ws/(2\pi)=10\,$Hz (below $\omegac$) the distribution shows no condensate peak, while for $\ws/(2\pi)=25\,$Hz (above $\omegac$) the lowest mode remains macroscopically occupied.

In the main paper we always use $\ws<\omegac$, which ensures that our drive uniformly mixes states with $k_z<\kc$.

\subsection{\textsc{III.~Extraction of the scaling exponents}}

\begin{figure}[t!]
\centerline{\includegraphics[width=\columnwidth]{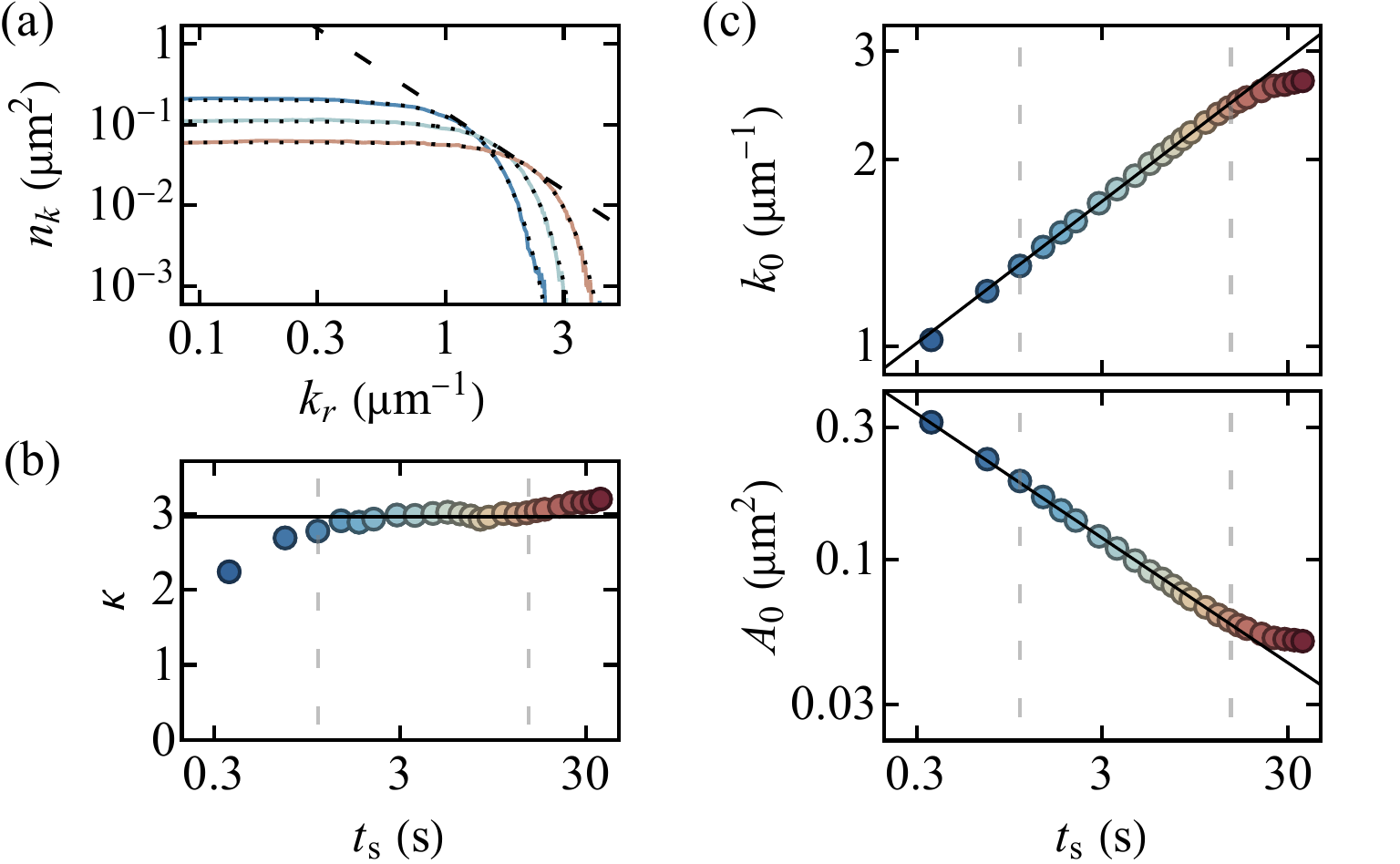}}
\caption{
Extraction of the scaling exponents; here $\Us/\kB=7.0$\,nK, $\ws/(2\pi)=10$\,Hz, and $\Gammad=8.0\,{\rm s}^{-1}$ (as in Fig.~4 of the main paper).  (a) Measured $n_k$ for $\ts=1.1$, $3.6$, and $14.4$\,s (solid lines) together with compressed-exponential fits (dotted lines). The dashed line is $\propto k_r^{-2}$, as in Fig.~4(b). (b) Fitted $\kappa(\ts)$. The dashed lines delineate the scaling range where $\kappa$ plateaus at $3.0(2)$ (solid line).
(c) Fixing $\kappa=3$, we refit $n_k$ curves to extract their widths $k_0(\ts)$ and amplitudes $A_0(\ts)$, and obtain $(\alpha,\beta)$ from the power-law fits $k_0\propto \ts^{-\beta}$ and $A_0\propto \ts^{\alpha}$~(solid lines).
}
\label{figS3}
\end{figure}

Here we detail our extraction of the scaling exponents $(\alpha,\beta)$ from $n_k(k_r, \ts)$.
We start by fitting $n_k(k_r)$ for different $\ts$ with $f_{\rm ce}=A_0 \exp[-(k_r/k_0)^\kappa]$, with $A_0$, $k_0$ and $\kappa$ as free parameters, as shown in Fig.~\ref{figS3}(a) for three experimental curves from Fig.~4(b) in the main paper.
As shown in Fig.~\ref{figS3}(b), we get essentially constant $\kappa$ for some $\ts$ range ($1.1-14.4$\,s); this is the scaling range in which the $n_k$ curves are self-similar.
We then fix $\kappa$ to its average value in the scaling range and refit the curves to obtain $k_0$ and $A_0$. Finally, as shown in Fig.~\ref{figS3}(c), we extract $\alpha$ and $\beta$ using power-law fits of the form $k_0\propto \ts^{-\beta}$ and $A_0\propto \ts^{\alpha}$~(solid lines).

\subsection{\textsc{IV.~Robustness of dynamic scaling}}
\begin{figure*}[hb!]
\centerline{\includegraphics[width=\textwidth]{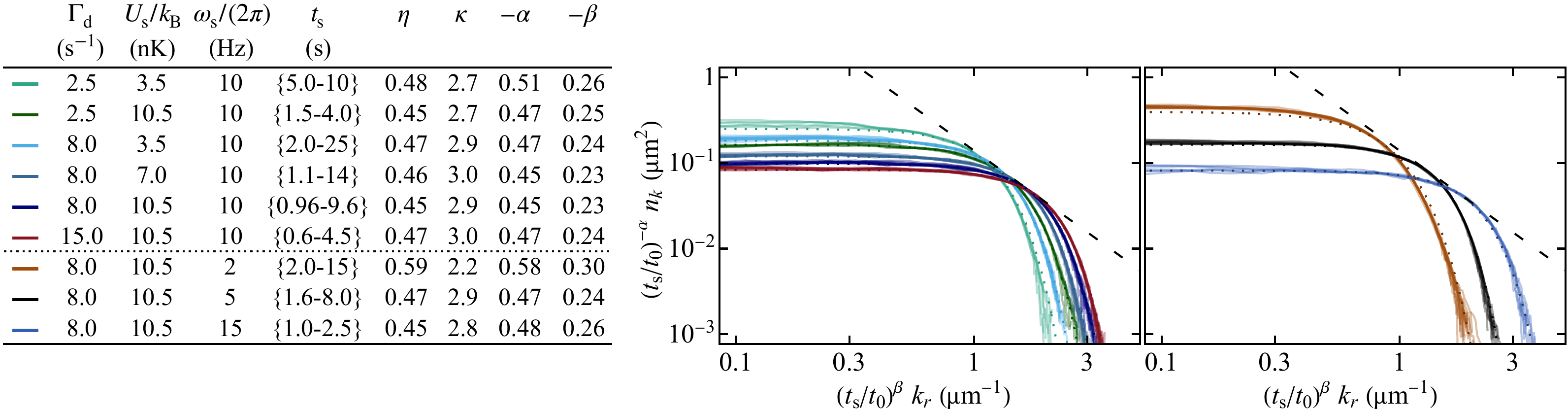}}
\caption{
Robustness of dynamic scaling. The table shows the extracted $\eta$, $\kappa$, $\alpha$, and $\beta$ values within the scaling ranges of $\ts$ for different $\Gammad$, $\Us$, and $\ws$.
The plots show the scaled $n_k$ with $t_0=3\,$s, for fixed $\ws$ and varying $\Gammad$ and $\Us$ (left panel), and for fixed $\Gammad$ and $\Us$ and varying $\ws$ (right panel). The uncertainties on $\beta$, $\alpha/2$, $\eta/2$ are typically $0.01$, which include uncertainties in determining the scaling range.
}
\label{figS4}
\end{figure*}

In Fig.~\ref{figS4} we show collapsed $n_k(\kr,\ts)$ for different disorder and drive parameters ($\Gammad$, $\Us$, and $\ws$), keeping the same reference time $t_0=3\,$s.
For fixed $\ws/(2\pi) = 10$\,Hz (left panel), over a factor of $6$ in $\Gammad$ and $3$ in $\Us$, we observe dynamic scaling with the same $\alpha/2\approx\beta =-0.24(2)$ and $\kappa=2.9(2)$. The collapsed curves are offset from each other because the dynamics are faster for larger $\Us$ or $\Gammad$ ($n_k$ propagates further in the same time $t_0$). 
Varying $\ws$ (right panel), as long as it is notably larger than $\omega_z = 2\pi \times 1.5$\,Hz and smaller than $\omegac$, we observe the same dynamic scaling behavior.
Curiously, we also observe subdiffusive scaling dynamics for $\ws \approx \omega_z$, but with notably different $\alpha/2\approx\beta=-0.30(1)$ and the scaling function with $\kappa=2.2(1)$, closer to a Gaussian.
\vspace{8em}

\onecolumngrid
~

\twocolumngrid

\end{document}